\documentclass[12pt]{article}
\usepackage{amsmath}
\usepackage{amssymb}
\usepackage{amsfonts}
\usepackage{latexsym}
\usepackage{color}
\usepackage{graphicx}

\catcode `\@=11 \@addtoreset{equation}{section}
\def\theequation{\arabic{section}.\arabic{equation}}
\catcode `\@=12

\renewcommand{\thesection}{\Roman{section}}

\newcommand{\be}{\begin{equation}}
\newcommand{\en}{\end{equation}}
\newcommand{\bea}{\begin{eqnarray}}
\newcommand{\ena}{\end{eqnarray}}
\newcommand{\beano}{\begin{eqnarray*}}
\newcommand{\enano}{\end{eqnarray*}}
\newcommand{\bee}{\begin{enumerate}}
\newcommand{\ene}{\end{enumerate}}

\newcommand{\Hil}{{\cal H}}

\newcommand{\F}{{\cal F}}
\newcommand{\Lc}{{\cal L}}

\newcommand{\G}{{\cal G}}

\newcommand{\E}{{\cal E}}

\newcommand{\1}{1 \!\! 1}

\begin{document}
\thispagestyle{empty}

\begin{center}
{\Large \bf Bicoherent-State Path Integral Quantization of a non-Hermitian Hamiltonian}   \vspace{1cm}\\

{\large F. Bagarello}
\vspace{3mm}\\[0pt]
Dipartimento di Ingegneria, Scuola Politecnica,\\[0pt]
Universit\`{a} di Palermo, I - 90128 Palermo,  and\\
\, INFN, Sezione di Napoli, Italy\\[0pt]
E-mail: fabio.bagarello@unipa.it\\[0pt]
home page: www1.unipa.it/fabio.bagarello \vspace{8mm}\\[0pt]

\vspace{0.4cm}

{\large Joshua Feinberg}
\vspace{3mm}\\[0pt]
Department of Mathematics and\\[0pt]
Haifa Research Center for Theoretical Physics and Astrophysics\\[0pt]
University of Haifa, Haifa 31905, Israel\\[0pt]
E-mail: jfeinberg@univ.haifa.ac.il\\[0pt]
https://orcid.org/0000-0002-2869-0010
\end{center}

\vspace*{0.5cm}

\begin{abstract}
We introduce, for the first time, bicoherent-state path integration as a method for quantizing non-hermitian systems. Bicoherent-state path integrals arise as a natural generalization of ordinary coherent-state path integrals, familiar from hermitian quantum physics. We do all this by working out a concrete example, namely, computation of the propagator of a certain quasi-hermitian variant of Swanson's model, which is not invariant under conventional $PT$-transformation. The resulting propagator coincides with that of the propagator of the standard harmonic oscillator, which is isospectral with the model under consideration by virtue of a similarity transformation relating the corresponding hamiltonians. We also compute the propagator of this model in position space by means of Feynman path integration and verify the consistency of the two results. \end{abstract}
\vspace{1cm}


\vfill

\newpage

\section{Introduction}\label{Introduction}

In this paper we focus on the quantum-mechanical oscillator described by the non-self-adjoint hamiltonian
\be
H_\theta=\frac{1}{2}\left(p^2+x^2\right)-\frac{i}{2}\,(\tan 2\theta)\left(p^2-x^2\right)\,. \label{200}
\en 
acting on wave-functions in the standard Hilbert space ${\cal H} =\Lc^2(\mathbb{R}).$ Here $\theta$ is a real parameter restricted to the range 
\be\label{range}
-\frac{\pi}{4}<\theta<\frac{\pi}{4}
\en
and $x$ and $p$ are the usual (hermitian) canonical position and momentum operators satisfying\footnote{We work in units in which $\hbar=1$.}\,\footnote{In order to avoid obfuscation of the physical aspects of our presentation, we shall ignore mathematical subtleties associated with unbounded operators, such as $x$, $p$ and $H_\theta$, except for occasions on which it is absolutely necessary to spell them out.}  $[x,p]=i\1$. 

This hamiltonian is a variant  \cite{dapro} of Swanson's model \cite{swan},  which unlike the latter, is {\em not} invariant under conventional $PT$-symmetry\cite{sense}. In fact, under $PT$-transformation, $H_\theta\rightarrow H_{-\theta} = H^\dagger_\theta$. Note, however, that the canonical transformation $x\rightarrow p\,, p\rightarrow -x,$ transforms $H_{-\theta}$ to $H_\theta$. Thus, $H_\theta$ and $H_{-\theta}$ are isospectral. 

We can rewrite $H_\theta$ in yet a different form as
\be
H_\theta=\frac{1}{2\cos 2\theta}\left[\left(e^{-i\theta}p\right)^2+\left(e^{i\theta}x\right)^2\right]. \label{Htheta}
\en  
Thus, $H_\theta$ is obtained from the standard harmonic oscillator $H_{\theta=0}$ (with spectrum $E_n=n+\frac{1}{2}$) by a complex canonical transformation
\be\label{canonical}
x\rightarrow X_\theta = e^{i\theta} x\,,\quad p\rightarrow P_\theta = e^{-i\theta}p\,,
\en 
such that 
\be\label{CCR}
[X_\theta,P_\theta] = i\1\,,
\en
followed by multiplication by an overall scale factor 
\be\label{omega}\omega_\theta=\frac{1}{\cos 2\theta}\,.\en
(Note that $\omega_\theta$ is well defined for all $\theta$ in the range (\ref{range}).) We thus conclude that the eigenvalues of $H_\theta$ are  
\be\label{spectrum}
E_n = \omega_\theta\left(n+\frac{1}{2}\right)\quad n= 0,1,2,\ldots\,.
\en 
This spectrum is an even function of $\theta$. Consequently, once again, we see that the {\em spectrum} of $H_\theta$ is invariant under $PT$-transformation. 
It obviously coincides with the spectrum of the conventional hermitian harmonic oscillator
\be\label{ho}
h_\theta = \omega_\theta\left(a^\dagger a + \frac{1}{2}\1\right),
\en
where $a = \frac{1}{\sqrt 2}(x+ip)$ and  $a^\dagger = \frac{1}{\sqrt 2}(x-ip)$ are the standard annihilation and creation operators. In fact, $H_\theta$ and $h_\theta$ are isospectral because they are related by the similarity transformation \cite{bagpl2010} 
\be\label{similarity}
H_\theta = T_\theta h_\theta T^{-1}_\theta\,,
\en
where 
\be\label{squeeze}
T_\theta = e^{\frac{i\theta}{2}(a^2-a^{\dagger 2})}
\en
is an unbounded, self-adjoint positive-definite operator, which we readily recognize as the squeezing operator \cite{Schumaker} with {\em imaginary} squeezing parameter $i\theta$. Consistency of \eqref{Htheta} and \eqref{similarity} then requires that 
\be\label{similarityxp}
X_\theta = T_\theta x T^{-1}_\theta \quad {\rm and}\quad P_\theta = T_\theta p T^{-1}_\theta \,,
\en
as well. This is indeed the case, because (\ref{canonical}) is nothing but the result of the non-unitary squeezing transformation with $T_\theta$. 

For the benefit of readers not familiar with squeezed coherent states, we note that in the position representation 
\be\label{squeeze-x}
T_\theta  = e^{-\frac{\theta}{2}(xp+px)} = e^{i\frac{\theta}{2}}e^{i\theta \frac{d}{d\log x}}\,.
\en
Thus, $T_\theta$ acting on any sensible function $\psi(x)$ shifts $\log x$ by $i\theta$, followed by a multiplication by an overall phase: $T_\theta \psi(x) = e^{i\frac{\theta}{2}}\psi(e^{i\theta}x)$. Using this representation of $T_\theta$ and the fact that $T_\theta^{-1} = T_{-\theta}$, we can prove that $T_\theta \left(x \left(T^{-1}_\theta\psi(x)\right)\right) = e^{i\theta}x\psi(x)$ in a straightforward manner. In a similar way, we can prove that $P_\theta\psi(x) = e^{-i\theta}p\psi(x) = -ie^{-i\theta}\frac{d\psi(x)}{dx}$.

The similarity transformation \eqref{similarityxp} implies that $X_\theta$ and $P_\theta$ are isospectral to $x$ and $p$. That is, $X_\theta$ and $P_\theta$ are diagonalizable with purely real spectra.  At the same time, \eqref{canonical} tells us that they are also proportional to $x$ and $p$, up to complex phases. However, this does not lead to a contradiction, because the similarity transformation \eqref{similarityxp} maps $\Hil$ onto $\Hil_\theta$. The relations \eqref{canonical} are only valid in $\Hil_\theta$, where $x$ and $p$ are {\em not} hermitian. On the other hand, the operators $x$ and $p$ which appear on the right-hand sides of the two equations in \eqref{similarityxp}, do operate on $\Hil$, where they are hermitian.

It follows from (\ref{similarity}) that $H^\dagger_\theta = T^{-1}_\theta h_\theta T\theta = T^{-2}_\theta H_\theta T^2_\theta\,.$ That is, $H^\dagger_\theta$ and $H_\theta $ are also related by a similarity transformation, or equivalently, satisfy the intertwining relation\footnote{Since $H_\theta$ and  $g_\theta$ are unbounded, this equation must be given a proper meaning, for instance by acting on vectors in some dense domain. Moreover, unbounded similarity transformations should be treated with extra care, since some eigenvectors of the transformed operator (the analog of $H_\theta$) may not belong to the domain of the transforming operator (the analog of $T_\theta$). Under such circumstances, the transformed operator and its resulting image (the analog of $H_\theta^\dagger$) are not isospectral. Happily, this is not the case in \eqref{similarity}.}  
\be\label{intertwining}
H^\dagger_\theta \,g_\theta = g_\theta H_\theta\,, 
\en
with 
\be\label{metric}
g_\theta =  T^{-2}_\theta = e^{i\theta(a^{\dagger 2} - a^2)} = e^{\theta(xp+px)}\,.
\en
The intertwining relation \eqref{intertwining} means that $H_\theta$ is actually hermitian with respect to the metric $g_\theta$, namely, in the Hilbert space ${\cal H}_\theta$ equipped with inner product 
\be\label{inner-product}
\langle \psi_1|\psi_2\rangle_\theta = \langle \psi_1|g_\theta|\psi_2\rangle\,.
\en
That is, 
\be\label{hermiticity}
\langle H_\theta\psi_1|\psi_2\rangle_\theta = \langle \psi_1|H_\theta\psi_2\rangle_\theta,
\en
by virtue of \eqref{intertwining}. As $\theta\rightarrow 0$, this Hilbert space turns, of course, into the standard Hilbert space ${\cal H}_{\theta=0}\equiv {\cal H} \,(\equiv\Lc^2(\mathbb{R}))$ with metric $g_0=\1$.

Clearly, $H_\theta$ is an observable in our theory, as are $X_\theta$ and $P_\theta$. More generally, any operator $A$ in this theory, which is hermitian with respect to $g_\theta$, that is, satisfies the intertwining relation 
\be\label{A-intertwining}
A^\dagger \,g_\theta = g_\theta A\,,
\en
is an observable. Such observables are sometimes referred to as a {\em quasi-hermitian} operators\cite{stellenbosch}, because $A$ is related to a hermitian operator by a similarity transformation, in a manner analogous to \eqref{similarity} and \eqref{similarityxp}.

The Swanson model \cite{swan} mentioned above, namely, the ${\cal PT}-$symmetric relative of (\ref{200}), offers yet another example of a quasi-hermitian hamiltonian. This well-studied model also demonstrates the fact that given the hamiltonian $H$, the metric $g$ is not unique \cite{swan, stellenbosch, Jones, Musumbu}. Various choices of $g$ lead to different quantizations of the system, with different irreducible sets of observables, in different Hilbert spaces.

More generally, the hamiltonian of any ${\cal PT}-$symmetric quantum mechanical system, with {\em unbroken}  ${\cal PT}$ symmetry, is hermitian with respect to the ${\cal CPT}$-inner product of that system \cite{sense}, and therefore possesses real spectrum. 

\subsection{Quantum dynamics}\label{dynamics}
Path integration is a method for computing matrix elements of the time evolution operator (or propagator) $e^{-iH_\theta t}$, which governs dynamics of the system. Thus, a few words are in order concerning dynamics of observables in our model. 

Time evolution in the Hilbert space ${\cal H}_\theta$ is unitary, namely, 
\be\label{unitarity}
e^{iH_\theta^\dagger t} g_\theta e^{-iH_\theta t}= g_\theta\,,
\en
which is a trivial consequence of \eqref{intertwining}.  

Consequently, Quantum Mechanics formulated in ${\cal H}_\theta$ is essentially not different from conventional hermitian Quantum Mechanics. To start with, the inner product of any two states  $|\psi_{1,2}(t)\rangle = e^{-iH_\theta t} |\psi_{1,2}(0)\rangle$, which evolve in time according to the Schr\"odinger equation, is time-independent: $\langle \psi_1(t)|\psi_2(t)\rangle_\theta = \langle \psi_1(0)|\psi_2(0)\rangle_\theta$. This implies, of course, probability conservation in the case of identical states.

By forming matrix elements of an observable $A=A(0)$ in the Schr\"odinger picture, we immediately deduce from \eqref{A-intertwining} that its Heisenberg picture counterpart $A(t)$ evolves according to 
\be\label{Heisenberg}
A(t)   = e^{iH_\theta t} A(0)e^{-iH_\theta t}\,.
\en
Equivalently, the Heisenberg equation of motion resulting from (\ref{Heisenberg}) is 
\be\label{Heisenbergeq}
i\dot A(t)  = [A(t),H_\theta]\,.
\en
Thus, observables which commute with $H_\theta$ are conserved in time.

It is trivial to verify that $A(t)$ fulfils the intertwining relation \eqref{A-intertwining} at all times, and is therefore an observable. This is consistent with the fact that the spectrum of $A(t)$ is conserved in time, because formally, \eqref{Heisenberg} is a similarity transformation\footnote{We qualify this similarity as {\em formal} because $e^{\pm iH_\theta t}$ are unbounded, in principle, and the range of, say, $e^{-iH_\theta t}$ needs not to be in the domain $D(A(0))$ of $A(0)$.}. 

\subsubsection{The quasi-hermitian position and momentum operators $X_\theta$ and $P_\theta$}\label{nonhermitianxp}
The Heisenberg equations of motion  \eqref{Heisenbergeq} for $X_\theta$ and $P_\theta$,
\be\label{HeisenbergXP}
\dot X_\theta = \omega_\theta P_\theta\quad{\rm and}\quad \dot P_\theta = -\omega_\theta X_\theta
\en 
are similar in form to the corresponding equations of motion in the hermitian problem. We can readily solve them and find 
\begin{eqnarray}\label{xp}
X_\theta(t) &=&  e^{iH_\theta t} X_\theta(0)e^{-iH_\theta t} = X_\theta(0) \cos\omega_\theta t + P_\theta(0)\sin\omega_\theta t\nonumber\\
P_\theta(t) &=&  e^{iH_\theta t} P_\theta(0)e^{-iH_\theta t} = P_\theta(0) \cos\omega_\theta t - X_\theta(0)\sin\omega_\theta t\,.
\end{eqnarray}
These solutions preserve the equal-time canonical commutation relation 
\be\label{ccrt}
[X_\theta(t),P_\theta(t)] = i\1
\en
at all times. 

Formally, by invoking the similarity transformations \eqref{similarity} and \eqref{similarityxp} to the middle term in each of the equations in \eqref{xp}, we can easily show that 
\begin{eqnarray}\label{xp1}
X_\theta(t) &=&  T_\theta \left(e^{ih_\theta t} x(0)e^{-ih_\theta t}\right) T_\theta^{-1} = T_\theta x(t) T_\theta^{-1}\nonumber\\
P_\theta(t) &=&  T_\theta \left(e^{ih_\theta t} p(0)e^{-ih_\theta t}\right) T_\theta^{-1} = T_\theta p(t) T_\theta^{-1}\,.
\end{eqnarray}
Thus, \eqref{similarityxp} holds for operators in the Heisenberg representation as well. $X_\theta(t)$ and $P_\theta(t)$ are (formally) similar to their hermitian counterparts $x(t)$ and $p(t)$. They are therefore diagonalizable, and their spectra are real as well. 

We shall now derive the spectral decompositions of $X_\theta(t)$ and $P_\theta(t)$ in terms of complete biorthogonal bases of right- and left-eigenvectors. To this end, we start by substituting the standard spectral decompositions 
\begin{eqnarray}\label{spectral-decomposition}
x(0) &=& \int\limits_{-\infty}^{\infty}dx\, x |x\rangle \langle x|\,,\quad \int\limits_{-\infty}^{\infty}dx\, |x\rangle \langle x|  = \1\,,\quad \langle x| x'\rangle = \delta (x-x')
\nonumber\\
p(0) &=& \int\limits_{-\infty}^{\infty}dp\, p |p\rangle \langle p|\,,\quad \int\limits_{-\infty}^{\infty}dp\, |p\rangle \langle p|  = \1\,,\quad \langle p| p'\rangle = \delta (p-p')
\end{eqnarray}
of $x(0)$ and $p(0)$, valid in the Hilbert space ${\cal H}$, in \eqref{xp1}. Therefore, 
\begin{eqnarray}\label{xp2}
X_\theta(t) &=&   \int\limits_{-\infty}^{\infty}dx\, x \left(T_\theta e^{ih_\theta t}  |x\rangle\right)\left( \langle x|  e^{-ih_\theta t} T_\theta^{-1} \right)  \nonumber\\
P_\theta(t) &=&  \int\limits_{-\infty}^{\infty}dp\, p \left(T_\theta e^{ih_\theta t}  |p\rangle\right)\left( \langle p|  e^{-ih_\theta t} T_\theta^{-1} \right)\,.
\end{eqnarray}
From this equation we can read-off the desired spectral decompositions in terms of complete biorthogonal bases of right- and left-eigenvectors for the time-dependent operators as follows. For $X_\theta(t)$ we obtain 
\begin{eqnarray}\label{tx}
X_\theta(t) &=& \int\limits_{-\infty}^{\infty}dx\, x |x;t\rangle_{\!R}\,  {}_{L}\!\langle x;t| \nonumber\\
|x;t\rangle_R &=& T_\theta |x;t\rangle  = T_\theta e^{ih_\theta t}  |x\rangle = e^{iH_\theta t}T_\theta|x\rangle\nonumber\\
|x;t\rangle_L &=& T_\theta^{-1}  |x;t\rangle = T_\theta^{-1} e^{ih_\theta t}  |x\rangle=g_\theta e^{iH_\theta t}T_\theta |x\rangle = g_\theta |x;t\rangle_R\,,
\end{eqnarray}
where we used \eqref{similarity} and \eqref{metric}, and invoked hermiticity of \eqref{squeeze}. Completeness and biorthogonality follow immediately: 
\be\label{tx-completenessorthogonal}
\int\limits_{-\infty}^{\infty}dx\, |x;t\rangle_{\!R}\,  {}_{L}\!\langle x;t|  = \1\,,\quad {}_{L}\!\langle x;t| x';t\rangle_{\!R} = \delta (x-x')\,.
\en
Similarly, for $P_\theta(t)$ we obtain 
\begin{eqnarray}\label{tp}
P_\theta(t) &=& T_\theta |p;t\rangle = \int\limits_{-\infty}^{\infty}dp\, p |p;t\rangle_{\!R}\,  {}_{L}\!\langle p;t| \nonumber\\
|p;t\rangle_R &=& T_\theta e^{ih_\theta t}  |p\rangle = e^{iH_\theta t}T_\theta|p\rangle\nonumber\\
|p;t\rangle_L &=& T_\theta^{-1}  |p;t\rangle = T_\theta^{-1} e^{ih_\theta t}  |p\rangle=g_\theta e^{iH_\theta t}T_\theta |p\rangle  = g_\theta |p;t\rangle_R\,,
\end{eqnarray}
and 
\be\label{tp-completenessorthogonal}
\int\limits_{-\infty}^{\infty}dp\, |p;t\rangle_{\!R}\,  {}_{L}\!\langle p;t|  = \1\,,\quad {}_{L}\!\langle p;t| p';t\rangle_{\!R} = \delta (p-p')\,.
\en
These spectral decompositions, into eigenstates with purely real eigenvalues $x$ and $p$, will be crucial for us in constructing the Feynman path integral for this system in Section V.

\subsubsection{The Classical Limit}\label{Classical limit}
The classical counterparts of the operators $X_\theta(t)$ and $P_\theta(t)$ in \eqref{xp} are real variables, corresponding to their parent quantum observables with their real spectra. This, together with the classical limit of \eqref{canonical}, means that $e^{i\theta}x$ and $e^{-i\theta}p$ should be taken as {\em real} variables as well.  That is, in the classical limit of the system originally defined in $\Hil_\theta$, the particle actually moves along the line in the complex $x$-plane making an angle $-\theta$ with the real axis. (These are precisely the anti-Stokes  lines discussed in Section \ref{stokes} below.) 

The classical variables $X_\theta(t)$ and $P_\theta(t)$obviously have canonical Poisson brackets, and any function $A(X_\theta(t), P_\theta(t))$ of these phase-space variables satisfies the canonical Hamiltonian equation of motion
\be\label{classical}
\dot A  = \{A, H_\theta\} = {\partial A\over\partial X_\theta}{\partial H_\theta\over\partial P_\theta} - {\partial A\over\partial P_\theta}{\partial H_\theta\over\partial X_\theta}\,,
\en
obtained from \eqref{Heisenbergeq} in the classical limit.

\subsection{Objectives}\label{objectives}
The main objective of this paper is to compute the propagator associated with (\ref{200}) by means of path integration based on bicoherent states \cite{bag2017}. Bicoherent states are a powerful tool in studying non-hermitian systems such as  (\ref{200}), and the present paper is the first application of bicoherent states to path integration. We shall also compute the transition amplitude of (\ref{200}) in position space, by means of Feynman path integration, and verify the consistency of the two results. As a byproduct of this computation, we shall gain insight into what Feynman path integrals of non-hermitian quantum systems really mean. 

\subsubsection{$\Hil_\theta$ vs. $\Hil$}\label{spaces}
The object of main interest in this paper is the probability amplitude 
\be\label{probability-amplitude}
A_{fi}(t) = \langle \psi_f |e^{-iH_\theta t}|\psi_i\rangle_\theta = \langle \psi_f |g_\theta e^{-iH_\theta t}|\psi_i\rangle
\en
for {\em unitary} time evolution of an initial state $|\psi_i\rangle$ into a final state $|\psi_f\rangle$. Time evolution is unitary, because \eqref{probability-amplitude} is defined in  the Hilbert space $\Hil_\theta$. In this evolution, the metric $g_\theta$ acts as a boundary term, at the end of the process. Its sole function is to ensure unitarity of the process, so that $|A_{fi}(t)|^2$ is the probability to start from  $|\psi_i\rangle$ and end up at  $|\psi_f\rangle$ after time $t$.  Now, imagine sandwiching the propagator $e^{-iH_\theta t}$ between two resolutions of unity associated with the position operator, as in the first equation in \eqref{spectral-decomposition}:
\begin{eqnarray}\label{probability-amplitude1}
A_{fi}(t) &=& \langle \psi_f |g_\theta \int\limits_{-\infty}^{\infty}dx_1\, |x_1\rangle \langle x_1| e^{-iH_\theta t} \int\limits_{-\infty}^{\infty}dx_2\, |x_2\rangle \langle x_2|\psi_i\rangle\nonumber\\ &=& \int\limits_{-\infty}^{\infty}dx_1\,dx_2\, \langle \psi_f |g_\theta |x_1\rangle \, \langle x_1| e^{-iH_\theta t} |x_2\rangle\,\langle x_2|\psi_i\rangle
\end{eqnarray}
The factors $\langle \psi_f |g_\theta |x_1\rangle $ and $\langle x_2|\psi_i\rangle$ are fixed position-dependent wave-functions, determined by the initial and final states. The interesting term, encoding the dynamics of our system, is the matrix element 
\be\label{propagator-matrix-element}
 G(x_1,x_2;t) = \langle x_1| e^{-iH_\theta t} |x_2\rangle
\en 
of the propagator. It is this object (or its analog, with $|x_{1,2}\rangle$ replaced by a pair of bicoherent states to be defined below) which we shall derive path-integral representations for. As it stands, we can think of it simply as a matrix element of a {\em non-hermitian} operator acting on the standard Hilbert space $\Hil$. This is the approach we shall adopt henceforth throughout the rest of this paper. 

In contrast to the hermitian case, \eqref{propagator-matrix-element} by itself is of course {\em not} a probability amplitude, but this should not prevent us from representing it as a path integral. After computing \eqref{propagator-matrix-element} by what-ever method we choose, we can plug it back into \eqref{probability-amplitude1} and compute the relevant probability amplitude.

\section{Pseudo-boson operator analysis of $H_\theta$}\label{sect2}
We use the standard annihilation and creation operators $a = \frac{1}{\sqrt 2}(x+ip)$, $a^\dagger = \frac{1}{\sqrt 2}(x-ip)$ to form the linear combinations \cite{bagpl2010}
\be\left\{
\begin{array}{ll}A_\theta= T_\theta  a T_\theta^{-1} =  a\cos\theta+ia^\dagger\sin\theta=\frac{1}{\sqrt{2}}\left(e^{i\theta}x+e^{-i\theta}\,\frac{d}{dx}\right),\\
B_\theta=T_\theta  a^\dagger T_\theta^{-1} =a^\dagger\cos\theta+ia\sin\theta=\frac{1}{\sqrt{2}}\left(e^{i\theta}x-e^{-i\theta}\,\frac{d}{dx}\right).\end{array} \right. \label{20}\en
It is clear that (for $\theta\neq 0$) $A_\theta^\dagger\neq B_\theta$ and also that
\be\label{com1} [A_\theta,B_\theta]=\1\,.\en Moreover, 
\be\label{com2}
[A_\theta,A^\dagger_\theta]= - [B_\theta,B^\dagger_\theta]=\1\cos 2\theta\,.\en  
For these reasons, we shall refer to $A_\theta, B_\theta$ and their hermitian adjoints as {\em pseudo-boson operators}. For further details and mathematical discussion of these operators, see \cite{bagpl2010,baginbagbook}.

We can use these pseudo-boson operators to write 
\be\label{Hpseudo}
H_\theta=\omega_\theta\left(B_\theta\,A_\theta+\frac{1}{2}\1\right)\,,\en

The two vacua of $A_\theta$ and $B_\theta^\dagger$ are defined, respectively, by 
\be\label{vacua1}\left\{
\begin{array}{ll}
A_\theta  \varphi_0^{(\theta)}(x) = T_\theta  a T_\theta^{-1}\varphi_0^{(\theta)}(x)=0\\{}\\
B_\theta^\dagger \Psi_0^{(\theta)}(x)= T_\theta^{-1}  a^\dagger T_\theta  \Psi_0^{(\theta)}(x)=0\,,
\end{array} \right.\en
where we have used hermiticity of $T_\theta$. Thus, $ \varphi_0^{(\theta)}(x) $ and $ \Psi_0^{(\theta)}(x)$ must be proportional, respectively, to $T_\theta$ and $T_\theta^{-1}$ acting on $\varphi_0(x)$ - the ground-state wave-function of the hermitian harmonic oscillator corresponding to $\theta=0$:
\begin{eqnarray}\label{vacua}
\varphi_0^{(\theta)}(x)&=& \alpha T_\theta  \varphi_0 (x)= N_\varphi
\exp\left(-\frac{1}{2}\,e^{2i\theta}\,x^2\right)\nonumber\\
\Psi_0^{(\theta)}(x)&=& \frac{1}{\alpha^*} T_\theta^{-1}\varphi_0 (x) = N_\Psi \exp\left(-\frac{1}{2}\,e^{-2i\theta}\,x^2\right)\,. \end{eqnarray}
The arbitrary complex proportionality constant $\alpha$ in the first equation determines that of the second equation as $\frac{1}{\alpha^*}$ due to the normalization condition 
\be\label{vacuum-normalization}
\left<\varphi_0^{(\theta)}\big|\Psi_0^{(\theta)}\right>=1
\en
(taken as  the usual $\Lc^2({\Bbb R})$ inner product of these states). This normalization condition must hold, because $\varphi_0^{(\theta)}(x)$ and $\Psi_0^{(\theta)}(x)$ comprise the first pair of left- and right-biorthogonal eigenvectors of $H_\theta$.  Equivalently, the normalization constants $N_\varphi$ and $N_\Psi$ in front of the Gaussian functions in \eqref{vacua} are constrained by \eqref{vacuum-normalization} according to 
\be\label{normalizationNN}
N^*_\varphi\,N_\Psi=\frac{e^{-i\theta}}{\sqrt{\pi}}\,.
\en
It is at this point that the reason for restricting the parameter $\theta$ to lie in the range (\ref{range}) becomes clear: In that range $\Re(e^{\pm 2i\theta})=\cos 2\theta>0$, and therefore both $\varphi_0^{(\theta)}(x)$ and $\Psi_0^{(\theta)}(x)$ (as well as all the other eigenstates, as can be seen from the discussion in the next subsection) belong to $\Lc^2({\Bbb R})$.

\subsection{Eigenstates}\label{eigensates} The eigenstates of $H_\theta$ and $H_\theta^\dagger$, respectively $\varphi_n^{(\theta)}(x)$ and $\Psi_n^{(\theta)}(x)$, can be derived in the usual manner \cite{bagpl2010},  by using the raising ladder operators $B_\theta$ and $A_\theta^\dagger$ :
\be
\left\{
\begin{array}{ll}
	\varphi_n^{(\theta)}(x)= \alpha T_\theta  \varphi_n (x)=\frac{1}{\sqrt{n!}}\,B_\theta^n\,\varphi_0^{(\theta)}(x)=\frac{N_\varphi}{\sqrt{2^n\,n!}}
	\,H_n\left(e^{i\theta}x\right)\,\exp\left(-\frac{1}{2}\,e^{2i\theta}\,x^2\right),  \\{}\\
	\Psi_n^{(\theta)}(x)=\frac{1}{\alpha^*} T_\theta^{-1}\varphi_n (x)=\frac{1}{\sqrt{n!}}\,(A_\theta^\dagger)^n\,\Psi_0^{(\theta)}(x)=\frac{N_\Psi}{\sqrt{2^n\,n!}}
	\,H_n\left(e^{-i\theta}x\right)\,\exp\left(-\frac{1}{2}\,e^{-2i\theta}\,x^2\right),
\end{array}
\right.
\label{21}\en
where $H_n(x)$ is the $n$-th Hermite polynomial. These are just the conventional eigenstates $\varphi_n(x)$ of the hermitian harmonic oscillator $H_{\theta=0}$, rotated into the complex-$x$ plane by $\pm\theta$. It then follows immediately from (\ref{com1}) (and its hermitian adjoint) and from (\ref{vacua1}) that 
\be\label{lowering}
A_\theta \varphi_n^{(\theta)}(x)= \sqrt{n}\varphi_{n-1}^{(\theta)}(x)\quad {\rm and}\quad
B_\theta^\dagger \Psi_n^{(\theta)}(x)=\sqrt{n}\Psi_{n-1}^{(\theta)}(x)\,,
\en
in complete analogy with the hermitian case.

The eigenstates (\ref{21}) define the two sets of functions,  $\F_\varphi^{(\theta)}=\{\varphi_n^{(\theta)}(x), \,n\geq0\}$ and $\F_\Psi^{(\theta)}=\{\Psi_n^{(\theta)}(x), \,n\geq0\}$. As discussed in detail in \cite{bagpl2010,baginbagbook}, these two sets are complete in $\Lc^2(\mathbb{R})$, but they are not bases. The proof of this claim is based on the fact that \cite{bagpl2010}
$$
\left\|\varphi_n^{(\theta)}\right\|^2=|N_\varphi|^2\sqrt{\frac{\pi}{\cos 2\theta}}P_n\left(\frac{1}{\cos 2\theta }\right),
$$
where $P_n(x)$ is the $n$-th Legendre Polynomial. Similarly, for $\left\|\Psi_n^{(\theta)}\right\|$, we have
$$
\left\|\Psi_n^{(\theta)}\right\|=\left|\frac{N_\Psi}{N_\varphi}\right|\left\|\varphi_n^{(-\theta)}\right\|=
\left|\frac{N_\Psi}{N_\varphi}\right|\left\|\varphi_n^{(\theta)}\right\|.
$$
Since $(\cos 2\theta )^{-1}>1$ one has \cite{szego} $P_n\left(\frac{1}{\cos 2\theta }\right)\rightarrow \infty$
as $n\rightarrow\infty$. Therefore both $\left\|\varphi_n^{(\theta)}\right\|$ and $\left\|\Psi_n^{(\theta)}\right\|$ diverge with $n$. Hence, see \cite{bagpl2010}, they cannot be bases for $\Lc^2(\mathbb{R})$.

Nevertheless, these two sets are biorthogonal, $\left<\varphi_n^{(\theta)}\big|\Psi_m^{(\theta)}\right>=\delta_{n,m}$, and they comprise $\Lc_\varphi$-{\em quasi-bases}. Here $\Lc_\varphi$ is the linear span of the functions $\varphi_n(x)=\varphi_n^{(0)}(x)=\Psi_n^{(0)}(x)$ (taken with $N_\varphi=N_\Psi=\frac{1}{\pi^{1/4}}$). The set $\{\varphi_n(x)\}$ is the orthonormal basis of the standard quantum harmonic oscillator, to which our model reduces at $\theta=0$. Hence the set $\Lc_\varphi$ is clearly dense in $\Lc^2(\mathbb{R})$. The fact that $\F_\varphi^{(\theta)}$ and $\F_\Psi^{(\theta)}$ are $\Lc_\varphi$-quasi bases means that, for all $f,g\in\Lc_\varphi$, 
\be
\left<f|g\right>=
\sum_n\left< f|\varphi_n^{(\theta)}\right>\left<\Psi_n^{(\theta)}|g\right>=\sum_n\left< f|\Psi_n^{(\theta)}\right>\left<\varphi_n^{(\theta)}|g\right>.
\label{22}\en
This equality is not necessarily {\em maximal}. That is to say, it might happen that a larger set $\G\supset\Lc_\varphi$ exists, such that (\ref{22}) can be extended to all $f,g\in\G$. We shall return to this issue later on. 

Many other details on the properties and applications of pseudo-bosons can be found in \cite{bagpl2010,baginbagbook}, to which we refer the interested reader. Here we are more interested in introducing the bicoherent states \cite{bag2017} for $H_\theta$, and use them to construct the bicoherent-state path integral of the model.

\subsection{Stokes wedges and observability of the operators $x$ and $p$}\label{stokes}
We have seen above that having $\theta$ lying in the range (\ref{range}) renders both $\varphi_0^{(\theta)}(x)$ and $\Psi_0^{(\theta)}(x)$ square-integrable along the real axis. In fact, all eigenstates decay at spatial infinity in both directions like Gaussians.  These states are vectors in the standard Hilbert space $\Hil = \Lc^2({\Bbb R})$ (recall the discussion in Section I.2.1), in which $x$ and $p$ are observables. Physically, the particle propagates along the real axis. Its position is a measurable quantity corresponding to the observable $x$. Similarly, its momentum is a measurable quantity corresponding to the observable $p$. 

It is instructive to reverse this logic, and see how the restriction (\ref{range}) on $\theta$ arises from the requirement that $x$ and $p$ be observables in $\Hil$, or equivalently, that $X_\theta$ and $P_\theta$ be observables in $\Hil_\theta$.  To this end, consider the standard Schr\"odinger eigenvalue equation of the harmonic oscillator (that is, $\theta = 0$ in (\ref{200}))
\be\label{schrodinger}
-\frac{1}{2}\psi''(x) + \frac{x^2}{2}\psi(x) = E\psi(x)\,,
\en
subjected to the boundary condition that $\psi (x)\rightarrow 0$ as $|x|\rightarrow\infty$. We can extend this problem to the complex-$x$ plane. Eigensolutions of (\ref{schrodinger}) vanish asymptotically in wedges of angular opening $\frac{\pi}{2}$ centered about the negative- and positive-real axes. These are the {\em Stokes wedges}  for this problem \cite{sense, BO, BB}. These wedges are bounded by the {\em Stokes lines} of the differential equation (\ref{schrodinger}). The real line, bisecting these wedges, is the {\em anti-Stokes line} of the problem. The eigensolutions of (\ref{schrodinger}) decay most rapidly along anti-Stokes lines and become purely oscillatory along Stokes lines. 

Let us now turn $\theta$ on. The eigenstates $\varphi_n^{(\theta)}(x)$ of $H_\theta$ are determined by solving (\ref{schrodinger}) but with $x\rightarrow X_\theta = e^{i\theta} x$, in accordance with (\ref{canonical}). The Stokes wedges for this problem are obtained by rotating the Stokes wedges of the original equation (\ref{schrodinger}) clockwise by angle $\theta$ when it is positive, or counter-clockwise by angle $|\theta|$ when it is negative. 

Recall the discussion in Section \ref{Classical limit} and note that the complex variable $X_\theta = e^{i\theta}x$  is real along the new anti-Stokes line. It would be interesting to investigate the relation between anti-Stokes lines in analytically continued spectral problems more generic than \eqref{schrodinger} and the range of the observable position (or momentum) operator, to see whether this is a generic feature. 

The spectral decomposition \eqref{spectral-decomposition} of $x$ in $\Hil$ and \eqref{tx} of $X_\theta$ in $\Hil_\theta$ involve integration over all real values of $x$, which we identify with the real axis in the complex $x$ plane associated with \eqref{schrodinger}. Thus, the real axis must remain within the Stokes wedges of the rotated problem. To this end, we have to restrict $|\theta|<\frac{\pi}{4}$, as in (\ref{range}).\\  
Solving the Schr\"odinger equation for the eigenstates $\Psi_n^{(\theta)}(x)$ of $H^\dagger_\theta$ leads to the same restriction on $\theta$, since $H^\dagger_\theta = H_{-\theta}$.

Again the same restriction (\ref{range}) on $\theta$ arises also from solving the Schr\"odinger eigenvalue equation for $H_\theta$ and $H_\theta^\dagger$ in the momentum representation, and demanding that $p$ be an observable in $\Hil$ and $P_\theta$ be an observable in $\Hil_\theta$, that is, that the real axis in the complex-$p$ plane always passes through the Stokes wedges as they are rotated according to $p\rightarrow P_\theta = e^{-i\theta}p$.

\subsection{The bicoherent states associated with $H_\theta$}\label{bico}

Bicoherent states were introduced in \cite{bag2017, bagbialo2017} as a generalization of conventional coherent states, pertaining to hermitian systems, to tackle non-hermitian quantum systems. In the present paper we focus exclusively on bicoherent states associated with the hamiltonian \footnote{For a general discussion of bicoherent states see \cite{bag2017, bagbialo2017}, and also \cite{baggargspa2018}.} (\ref{200}). These bicoherent states are built upon the two biorthogonal families of vectors $\F_\varphi^{(\theta)}$ and $\F_\Psi^{(\theta)},$ which are $\Lc_\varphi$-quasi bases, as indicated by (\ref{22}). Thus, based on (\ref{vacua1}) and (\ref{lowering}), and in complete analogy with the construction of conventional coherent states \cite{gazeaubook, klauder, schulman}, the desired bicoherent states are given in terms of the two series 
\begin{eqnarray}\label{bicoherent}
\varphi^{(\theta)}(z,x) &=& N(|z|)\sum_{k=0}^\infty {z^k\over\sqrt{k!}}\varphi_k^{(\theta)}(x)\nonumber\\
\Psi^{(\theta)}(z,x) &=& N(|z|)\sum_{k=0}^\infty {z^k\over\sqrt{k!}}\Psi_k^{(\theta)}(x)\,,
\end{eqnarray}
with a common $\theta$-independent normalization factor 
\be\label{normalizationN}
N(|z|)  = \left(\sum_{k=0}^\infty{ |z|^{2k}\over k!}\right)^{-\frac{1}{2}} = e^{-\frac{1}{2}|z|^2}\,.
\en
As is evident from the first equality in each of the equations \eqref{21}, the bicoherent states are related to the conventional coherent states 
\be\label{coherent-conventional}
\Phi(z,x)  = N(|z|)\sum_{k=0}^\infty {z^k\over\sqrt{k!}}\varphi_k (x)
\en
of the hermitian harmonic oscillator according to
\begin{eqnarray}\label{bicoherent-simple}
\varphi^{(\theta)}(z,x) &=& \alpha T_\theta \Phi(z,x) \nonumber\\
\Psi^{(\theta)}(z,x) &=& \frac{1}{\alpha^*} T_\theta^{-1}\Phi(z,x) \,, 
\end{eqnarray}
where $\alpha$ was defined following \eqref{vacua}.

Based on the detailed analysis made in \cite{bag2017, bagbialo2017}, the two series in (\ref{bicoherent}) can be shown to converge in the entire complex-$z$ plane, for all values of $\theta$ in the range (\ref{range}). Refs. \cite{bag2017, bagbialo2017} fell short of explicit expressions for these series as functions of $z$ and $x$, which we now provide:  From the definition (\ref{21}) of eigenstates, and from Rodrigues'  representation
\be\label{rodrigues}
H_n(u) = (-1)^n e^{u^2}{\partial^n\over\partial u^n} e^{-u^2}
\en
for Hermite's polynomials \cite{Abramowitz}, we can derive, after some straightforward steps, the following explicit expressions 
\begin{eqnarray}\label{bicoherent1}
\varphi^{(\theta)}(z,x) &=& N_\varphi \exp\left[-\frac{1}{2}\left(|z|^2 + z^2 + e^{2i\theta}x^2\right) + \sqrt{2} ze^{i\theta}x\right]\nonumber\\
\Psi^{(\theta)}(z,x) &=& N_\Psi\exp\left[-\frac{1}{2}\left(|z|^2 + z^2 + e^{-2i\theta}x^2\right) + \sqrt{2} ze^{-i\theta}x\right]\,,
\end{eqnarray}
for these states in the position representation.  In the limit $\theta\rightarrow 0$ (in which $T_\theta\rightarrow\1$), these expressions coincide with the conventional coherent-state wave functions in the position representation. (See {\it e.g.,} Eq. (27.12) in \cite{schulman}.) In other words, (\ref{bicoherent1}) are appropriate analytic continuations of the conventional coherent state position wave functions into the complex-$x$ plane, replacing $x$ with $e^{\pm i\theta}x$, in accordance with \eqref{canonical}.

A defining property of conventional coherent states is that they are eigenstates of the annihilation operator. Our bicoherent states enjoy a natural generalization of this attribute.  Indeed, we can use (\ref{lowering}) to show, in a straightforward manner, that $\varphi^{(\theta)}(z,x)$ and $\Psi^{(\theta)}(z,x)$ are eigenstates of the two annihilation operators at hand, namely, 
\be\label{bicoherent2}
A_\theta\varphi^{(\theta)}(z,x) = z\varphi^{(\theta)}(z,x)\quad{\rm and}\quad B_\theta^\dagger\Psi^{(\theta)}(z,x) = z\Psi^{(\theta)}(z,x)\,.
\en

Another defining property of conventional coherent states is that they resolve the identity. (In fact, they form an over-complete  set.) This is also an attribute of bicoherent states, which turn out to resolve (together) the identity, at least on the set $\Lc_\varphi$, which is dense in $\Lc^2(\mathbb{R})$.  To see this, we have to verify that\footnote{Here we have anticipated the measure $ {d^2z\over\pi} =\frac{1}{\pi} d{\rm Re}\,z\, d{\rm Im}\,z $ from our experience with conventional coherent states.}
\be\label{resolution}
\langle f| g\rangle = \int_{\mathbb{C}} {d^2z\over\pi} \langle f|\varphi^{(\theta)}(z)\rangle \langle \Psi^{(\theta)}(z) | g\rangle
\en
for any pair of vectors  $|f\rangle, |g\rangle \in \Lc_\varphi$. Integration over the complex plane in (\ref{resolution}) is straightforward, based on the definition (\ref{bicoherent}) and on the elementary integral 
$$ \int_{\mathbb{C}} {d^2z\over\pi} e^{-|z|^2} {z^m z^{*n}\over\sqrt{m! n!}} = \delta_{mn}\,.$$
The result is 
\be\label{resolution1}
\int_{\mathbb{C}} {d^2z\over\pi} \langle f |\varphi^{(\theta)}(z)\rangle \langle \Psi^{(\theta)}(z) | g\rangle =  \sum_n\left< f |\varphi_n^{(\theta)}\right>\left<\Psi_n^{(\theta)} |g\right>   = \langle f | g\rangle
\en
as required, by virtue of (\ref{22}).

In a similar manner, we can establish the alternative resolution of the identity 
\be\label{resolution-alternative}
\langle f |g\rangle = \int_{\mathbb{C}} {d^2z\over\pi} \langle f |\Psi^{(\theta)}(z)\rangle \langle \varphi^{(\theta)}(z)  | g\rangle\,.
\en

In complete analogy with conventional coherent states, the overlap matrix element of two bicoherent states has the {\em reporoducing kernel} property \cite{klauder}. 
More specifically, consider 
\be\label{overlap}
K(z_1,z_2) = \langle \Psi^{(\theta)}(z_1)| \varphi^{(\theta)}(z_2)\rangle
\en
for any pair $z_1,z_2\in\mathbb{C}$ of complex variables. This overlap is well defined for all $z_1,z_2\in\mathbb{C}$ due to the Schwarz inequality. In order to compute it, recall the definition (\ref{bicoherent}) and expand 
\begin{eqnarray}\label{Kexpansion}
K(z_1,z_2)  &=& e^{-\frac{1}{2}(|z_1|^2+|z_2|^2)}\sum_{k,l=0}^\infty \frac{z_1^{*k}\,z_2^l}{\sqrt{k!\,l!}}\left<\Psi_k^{(\theta)}\big|\varphi_l^{(\theta)}\right> \nonumber\\
&=&e^{-\frac{1}{2}(|z_1|^2+|z_2|^2)}\sum_{k=0}^\infty \frac{(z_1^*\,z_2)^k}{k!}=e^{-\frac{1}{2}(|z_1|^2+|z_2|^2)+ z_1^*\,z_2}\,.
\end{eqnarray}
This is manifestly $\theta-${\em independent}, and holds in particular for $\theta=0$, namely, for the overlap of coherent states of the standard harmonic oscillator 
$\Phi(z,x)=e^{-|z|^2/2}\sum_{k=0}^\infty\frac{z^k}{\sqrt{k!}}\,\varphi_k(x) = \Psi^{(0)}(z,x) = \varphi^{(0)}(z,x)$. $\theta$-independence of \eqref{Kexpansion} should come at no surprise, since due to \eqref{bicoherent-simple} we can write 
\be\label{overlap1}
K(z_1,z_2) = \Big\langle \frac{1}{\alpha^*} T_\theta^{-1} \Phi (z_1)\Big | \alpha T_\theta \Phi(z_2)\Big\rangle =  \left<\Phi(z_1)|\Phi(z_2)\right> \,,
\en
where we used hermiticity of $T_\theta$ in \eqref{squeeze}. 

Thus, $K(z_1,z_2)$ coincides with the reproducing kernel 
of conventional coherent states 
\be\label{reproducing}
K(z_1,z_2) = \left<\Phi(z_1)|\Phi(z_2)\right> = \langle \Psi^{(\theta)}(z_1)| \varphi^{(\theta)}(z_2)\rangle= e^{-\frac{1}{2}(|z_1|^2+|z_2|^2)+ z_1^*\,z_2}\,,
\en
and therefore satisfies the reproducing relation \cite{klauder}
\be\label{reproducing1}
 \int_{\mathbb{C}} {d^2z_2\over\pi}K(z_1,z_2) K(z_2,z_3) = K(z_1,z_3)\,.
\en
In a similar manner, we can also show that
\be\label{reproducing2}
K(z_1,z_2) =  \left<\varphi^{(\theta)}(z_1)|\Psi^{(\theta)}(z_2)\right>\,,
\en
which is equivalent to hermiticity of the inner product in (\ref{reproducing}), namely, $$K^*(z_1,z_2) =  \left<\varphi^{(\theta)}(z_2)|\Psi^{(\theta)}(z_1)\right> = K(z_2,z_1)\,.$$
Needless to say, $\langle \Psi^{(\theta_1)}(z_1)| \varphi^{(\theta_2)}(z_2)\rangle$, when $\theta_1\neq\theta_2$, is still well defined, but does not satisfy (\ref{reproducing1}).

Finally, note that (\ref{reproducing}) and (\ref{reproducing2}), together with (\ref{reproducing1}) imply that 
\be
K(z_1,z_2) = \langle \Psi^{(\theta)}(z_1)| \varphi^{(\theta)}(z_2)\rangle = \int_{\mathbb{C}} {d^2z\over\pi} \langle \Psi^{(\theta)}(z_1)|\varphi^{(\theta)}(z)\rangle \langle \Psi^{(\theta)}(z) | \varphi^{(\theta)}(z_2\rangle\,.
\label{213}\en
We have proved the resolution of unity (\ref{resolution}) by bicoherent states in its weakest form, restricting ourselves to the set $\Lc_\varphi$. We stress that (\ref{213}) extends this resolution well beyond $\Lc_\varphi$. This is so because (in comparison with (\ref{resolution})) the pair of bicoherent states  $|f\rangle = |\Psi^{(\theta)}(z_1)\rangle$ and $|g\rangle = |\varphi^{(\theta)}(z_2)\rangle$ manifestly do not belong in $\Lc_\varphi$ (they are not finite linear combinations of the 
$|\varphi_n\rangle$'s). Yet, (\ref{213}) demonstrates explicitly that the resolution of unity (\ref{resolution}) holds for such states as well. 

\newpage

\section{The propagation amplitude}\label{prop-amplitude}
 In this section we derive explicitly the matrix elements of the propagator $e^{-iH_\theta t}$ in position space and in terms of bicoherent states.
 
\subsection{The bicoherent propagation amplitude}\label{bicoherent-amplitude}

Consider the matrix element of the propagator between a pair of bicoherent states
 \be
 D^{({\theta})}(z_f,z_i;t)=\left<\Psi^{({\theta})}(z_f)| e^{-iH_\theta t}|\varphi^{({\theta})}(z_i)\right>,
 \label{33}\en
for initial and final complex parameters $z_i, z_f$. The matrix elements (\ref{33}), like their position-space counterparts  \eqref{propagator-matrix-element}, contain all the information about the propagator $e^{-iH_\theta t}$, due to (over-)completeness of bicoherent states, as was discussed in \ref{bico}. This renders $ D^{({\theta})}(z_f,z_i;t)$ an object of utmost importance for us. Indeed, by invoking 
the completeness relations (\ref{resolution}), (\ref{resolution-alternative}), or (\ref{213}) (depending on which subsets of Hilbert space $|\psi_i\rangle$ and $|\psi_f\rangle$ belong to), we can obtain (\ref{probability-amplitude}) from (\ref{33}) as 
 \be
 A_{fi}(t)=\int_{\mathbb{C}}{d^2z_f\over\pi}{d^2z_i\over\pi} \left<\psi_f|g_\theta
 \varphi^{(\theta)}(z_f)\right>D^{(\theta)}(z_f,z_i;t_f)\left<\Psi^{(\theta)}(z_i)| \psi_i\right>\,.
 \label{32}\en

We conclude this short subsection by computing (\ref{33}) explicitly. To this end, apply the propagator $e^{-iH_\theta t}$ to the bicoherent state $\varphi^{({\theta})}(z_i)$ in (\ref{bicoherent}), and recall from (\ref{21}) that $H_\theta \varphi_k^{({\theta})}(x) = \omega_\theta (k+\frac{1}{2})\varphi_k^{({\theta})}(x)$. Thus, 
\be
 e^{-iH_\theta t} \varphi^{({\theta})}(z_i)= e^{-i\omega_\theta t/2}\varphi^{({\theta})}\left(z_ie^{-i\omega_\theta t}\right)\,.
 \label{34}\en
We now plug this into (\ref{reproducing}) and obtain the desired result
\begin{eqnarray}
D^{({\theta})}(z_f,z_i;t)&=&e^{-i\omega_\theta t/2} \left<\Psi^{({\theta})}(z_f)|\varphi^{({\theta})}(z_ie^{-i\omega_\theta t})\right>\nonumber\\{}\nonumber\\
&=&e^{-i\omega_\theta t/2}\exp\left[-\frac{1}{2}(|z_f|^2+|z_i|^2)+z^*_f\,z_ie^{-i\omega_\theta t}\right]\,.
\label{35}\end{eqnarray}
An alternative derivation of \eqref{35} starts from \eqref{bicoherent-simple}, according to which 
\begin{eqnarray}
D^{({\theta})}(z_f,z_i;t)&=& \left(\frac{1}{\alpha^*} T_\theta^{-1} |\Phi(z_f)\rangle\right)^\dagger e^{-iH_\theta t}\alpha T_\theta|\Phi(z_i)\rangle
\nonumber\\{}\nonumber\\
&=&\langle\Phi(z_f)| T_\theta^{-1} e^{-iH_\theta t} T_\theta |\Phi (z_i)\rangle \nonumber\\{}\nonumber\\
&=& \langle\Phi(z_f)|  e^{-ih_\theta t}  |\Phi (z_i)\rangle \,,
\label{35a}\end{eqnarray}
where we have first used hermiticity of $T_\theta$ and then the similarity transformation \eqref{similarity}.
Thus, $D^{({\theta})}(z_f,z_i;t)$ coincides with the analogous coherent-state matrix element for the {\em hermitian} harmonic oscillator with frequency $\omega_\theta$, as is evident \cite{schulman} from the last explicit form equation in \eqref{35}.

\subsection{The propagator in position basis}\label{position-amplitude}
The propagator matrix element (\ref{propagator-matrix-element}) in position space
\be\label{meposition}
G(x_f,x_i;t)  = \langle x_f|e^{-iH_\theta t}|x_i\rangle
\en
can be computed by substituting $\langle x_f| \varphi^{(\theta)}(z_f)\rangle = \varphi^{(\theta)}(z_f,x_f)$ and $\langle\Psi^{(\theta)}(z_i)|x_i\rangle = (\Psi^{(\theta)}(z_i,x_i))^*$ from (\ref{bicoherent1}) in (\ref{32}), together with (\ref{35}) and the complex conjugate of (\ref{normalizationNN}). We end up with  
\be\label{meposition1}
G(x_f,x_i;t)  = {e^{i\theta}\over\sqrt{\pi}} e^{-\frac{i}{2}\omega_\theta t} e^{-\frac{1}{2}\left[(e^{i\theta} x_f)^2 + (e^{i\theta}x_i)^2\right]}I(x_f,x_i;t)\,
\en
where the double-Gaussian integral 
\be\label{Iint}
I(x_f,x_i;t) = \int_{\mathbb{C}}{d^2z_f d^2z_i\over\pi^2} \exp\left\{\!-|z_f|^2\!-|z_i|^2\!-\frac{1}{2}(z_f^2 + z_i^{*2})\! + \sqrt{2}e^{i\theta}(z_f x_f +z^*_i x_i)\!  +z_f^*z_ie^{-i\omega_\theta t}\right\}
\en
is computed in the Appendix. The final result is 
\be\label{meposition2}
G(x_f,x_i;t)  = {e^{i\theta}\over\sqrt{2\pi i \sin\omega_\theta t}}  \exp\left\{\frac{ie^{2i\theta}}{2\sin\omega_\theta t}\left[(x_f^2 + x_i^2)\cos\omega_\theta t - 2x_fx_i\right]\right\}\,.
\en
By comparing (\ref{meposition2}) with the analogous matrix element of the conventional hermitian oscillator $h_\theta = \omega_\theta\left(a^\dagger a + \frac{1}{2}\1\right)$
in \eqref{ho},
\be\label{meposition-hermitian}
 \langle x_f|e^{-ih_\theta t} |x_i\rangle  = {1\over\sqrt{2\pi i \sin\omega_\theta t}}  \exp\left\{\frac{i}{2\sin\omega_\theta t}\left[(x_f^2 + x_i^2)\cos\omega_\theta t - 2x_fx_i\right]\right\}\,.
\en
(see e.g. Eq. (6.38) in \cite{schulman}), and with \eqref{Htheta} and \eqref{canonical} in mind, it is gratifying to note that (\ref{meposition2}) is nothing but the former expression, analytically continued according to 
\be\label{ancont}
x_{i,f}\rightarrow e^{i\theta} x_{i,f}\,,
\en
followed by an overall multiplicative factor $e^{i\theta}$. (See Section \ref{Feynman} for more details.) 

Finally, as a trivial check, note that $G(x_f,x_i;t) \rightarrow\delta(x_f-x_i)$ (weakly) as $t\rightarrow 0$, in accordance with (\ref{meposition}). 

\newpage 
\section{The Bicoherent-State Path integral}\label{bicoherentpath}
The main objective of the present work is derivation of the bicoherent-state path integral representation for the propagator (\ref{35}). Surely enough, the path integral derivation of (\ref{35}) which follows is not as simple and straightforward as the direct derivation in Section \ref{bicoherent-amplitude}. For simple systems such as (\ref{200}), path integration techniques are evidently an over-kill. Moreover, since (\ref{35}) (and consequently (\ref{transfermatrix}) below) are identical to the analogous quantities for the conventional hermitian oscillator with frequency $\omega_\theta$, the bicoherent-state path integral we are about to derive in this section for the propagator of the nonhermitian oscillator (\ref{200}) will be identical with the conventional coherent-state path integral of the hermitian oscillator. Nevertheless, we shall pursue its derivation in what follows, because our purpose here is to take advantage of the simplicity of (\ref{200}) and use it to introduce bicoherent-path integration as a novel technique for quantizing more complicated {\em interacting} non-hermitian systems, including non-hermitian quantum field theories. Path integration may be the preferable method for quantizing such systems. Our purpose here is to introduce this method and demonstrate that it works. 

\subsection{Slicing the time axis}\label{slicingtime}
The standard first step in constructing the path integral is to slice the segment of the time axis between the initial $t_i=0$ and final $t_f=t >0$ times into small $N$ segments:
\be\label{slicing}
0=t_i=t_0 < t_1<t_2 <\cdots <t_{N-1}<t_N=t_f=t\,.
\en
With no loss of generality (and for simplicity), we shall take all these segments to be of equal duration 
\be\label{segment}
\epsilon=t_{k+1}-t_k\,,
\en
so that 
\be\label{duration}
t=N\epsilon\,.
\en
The total propagator  $e^{-iH_\theta t}$ is the result of propagation along the $N$ consecutive time segments. Thus, we write the bicoherent matrix element (\ref{33}) as 
\be\label{33a}
D^{({\theta})}(z_N,z_0;t)=\left<\Psi^{({\theta})}(z_N)\big | \left(e^{-iH_\theta \epsilon}\right)^N\big |\varphi^{({\theta})}(z_0)\right>,
\en 
where we have renamed $z_i=z_0$ and $z_f=z_N$, in accordance with notation introduced in (\ref{slicing}). Next, as in (\ref{213}), we insert $N-1$ resolutions of the identity at the intermediate slicing points in (\ref{33a}), labelling the bicoherent state complex parameter at time $t_k$ by $z_k$. Note from (\ref{34}) that 
\be\label{34a}
 e^{-iH_\theta \epsilon} \varphi^{({\theta})}(z_k)= e^{-i\omega_\theta \epsilon/2}\varphi^{({\theta})}\left(z_ke^{-i\omega_\theta \epsilon}\right)\,.
\en
By making use of this fact, and by invoking the reproducing kernel property (\ref{reproducing1}) repeatedly when integrating over the intermediate complex variables, we can rewrite $D^{({\theta})}(z_N,z_0;t)$ as 
 \be
 D^{({\theta})}(z_N,z_0;t)=\int \left[\prod_{k=1}^{N-1} \,{d^2 z_k\over\pi} \right]\,\prod_{l=1}^{N} I(z_l,z_{l-1}),
 \label{41}\en
 where 
 \be
 I(z_l,z_{l-1})=\left<\Psi^{({\theta})}(z_l)|e^{-iH_\theta\epsilon}|\varphi^{({\theta})}(z_{l-1})\right> = D^{({\theta})}(z_l,z_{l-1};\epsilon).
 \label{42}\en
 $I(z_l,z_{l-1})$ is sometimes referred to as an element of the {\em transfer matrix}, as it transfers, or propagates, the system from one time slice to the next one. 
 Thus, from (\ref{35}) we obtain
\be\label{transfermatrix}
 I(z_l,z_{l-1})=e^{-i\omega_\theta \epsilon/2}\exp\left\{-\frac{1}{2}(|z_l|^2+|z_{l-1}|^2)+z^*_l\,z_{l-1}e^{-i\omega_\theta \epsilon}\right\}.
\en
Substituting (\ref{transfermatrix})  in (\ref{41}) we conclude that
\be\label{Delement}
  D^{({\theta})}(z_N,z_0;t)=e^{-i\omega_\theta t/2}e^{-\frac{1}{2}(|z_N|^2+|z_{0}|^2)}\E(z_N,z_0),
 \en
 where
 \be\label{E}
 \E(z_N,z_0)=\int \left[\prod_{k=1}^{N-1} \,{e^{-|z_k|^2}d^2 z_k\over\pi} \right]\, e^{e^{-i\omega_\theta \epsilon}(z^*_N\,z_{N-1}+z^*_{N-1}\,z_{N-2}+\cdots+z^*_1\,z_{0})}.
 \en
Computation of $\E(z_N,z_0)$ is based on the simple integral 
 \be
 \int_{\mathbb{C}}\,{e^{-|z|^2}d^2 z\over\pi}\,e^{w^*_2\,z+z^*\,w_1}=e^{w^*_2\,w_1}\,.
 \label{43}\en
Hence, for instance,
 $$
 \int \,{e^{-|z_1|^2}d^2 z_1\over\pi}\,e^{e^{-i\omega_\theta \epsilon}(z^*_2\,z_{1}+z^*_1\,z_{0})}=e^{z^*_2\,z_{0}\,e^{-2i\omega_\theta\epsilon}},
 $$
  $$
 \int \,{e^{-|z_2|^2}d^2 z_2\over\pi}\,e^{e^{-i\omega_\theta \epsilon}z^*_3\,z_2+z^*_2\,z_0\,e^{-2i\omega_\theta\epsilon}}=e^{z^*_3\,z_0\,e^{-3i\omega_\theta\epsilon}},
 $$
 and so on, until the last integration over $z_{N-1}$ yields 
 \be\label{E1}
 \E(z_N,z_0)=e^{z^*_N\,z_0\,e^{-iN\omega_\theta\epsilon}}=e^{z^*_N\,z_0\,e^{-i\omega_\theta t}}\,.
 \en
Finally, by plugging this result back in (\ref{Delement}), we recover (\ref{35}):   
 \be
D^{({\theta})}(z_N,z_0;t)=e^{-i\omega_\theta t/2}e^{-\frac{1}{2}(|z_N|^2+|z_{0}|^2)}\E(z_N,z_0)=e^{-i\omega_\theta t/2}e^{-\frac{1}{2}(|z_N|^2+|z_{0}|^2)+z^*_N\,z_{0}\,e^{-i\omega_\theta t}}\,.
  \label{44}\en

\subsection{The limit of infinitely many time slices - the path integral} 
The discussion in the previous subsection is exact for any number $N$ of time slices. It is evident from this discussion that we always recover the correct matrix element (\ref{35}), whatever the number $N$ of time slices is. 

Let us think of the series of $N$ points $z_l$ as snapshots of a function $z(t)$ taken at the times $t_l$ in (\ref{slicing}):
\be\label{snapshot}
z_l=z(t_l)\,.
\en
As $N$ gets larger, the points $z_l$ sample the function $z(t)$ at an ever increasing resolution. Thus, in the limit $N\rightarrow\infty$, that is $\epsilon\rightarrow 0$, we recover the function $z(t)$, a trajectory in the complex plane,  in its entirety.  With this picture in mind, the expression (\ref{41}) for the propagation amplitude  $D^{({\theta})}(z_N,z_0;t)$ can be interpreted, in the limit $N\rightarrow\infty$, as a sum over all possible paths $z(t)$ in the complex plane, connecting $z = z_0$ at $t=0$ with $z=z_N$ at time $t$, where each path is weighed with a complex amplitude which we derive below. This is the bicoherent-state path integral representation of (\ref{33}). 

We proceed by rewriting (\ref{transfermatrix}) as 
\be\label{transfermatrix1}
I(z_l,z_{l-1})=e^{-i\omega_\theta \epsilon/2}\exp\left\{-\frac{1}{2}\left[z^*_l(z_l-z_{l-1}) - (z^*_l-z^*_{l-1})z_{l-1}\right]+z^*_l\,z_{l-1}\left(e^{-i\omega_\theta \epsilon}-1\right)\right\}.
\en
It is at this point that we start making approximations in the limit $\epsilon\rightarrow 0$. Evidently,  one can interpret the differences appearing in the second exponential in (\ref{transfermatrix1}) as time-derivatives of $z(t)$,
\begin{equation}\label{time-derivative}
z_{l+1} - z_l \simeq \epsilon \dot z_{l+1}\,, 
\end{equation}
so that, up to corrections of ${\cal O}(\epsilon^2)$, we can write
\be\label{transfermatrix2}
I(z_l,z_{l-1})=e^{-i\omega_\theta \epsilon/2}\exp\left\{-\frac{\epsilon}{2}\left[z^*_l\dot z_l - \dot z^*_l(z_l-\epsilon\dot z_l) \right]+z^*_l(z_l-\epsilon\dot z_l)\left(e^{-i\omega_\theta \epsilon}-1\right)\right\},
\en
or, more explictly, 
\be\label{transfermatrix3}
I(z_l,z_{l-1})=e^{-i\omega_\theta \epsilon/2}\exp\left\{-\frac{\epsilon}{2}\left(z^*_l\dot z_l - \dot z^*_lz_l \right) -i\epsilon\omega_\theta z^*_lz_l +{\cal O}(\epsilon^2)\right\},
\en
for the transfer matrix element. We finally plug (\ref{transfermatrix3}) in (\ref{41}), neglect the ${\cal O}(\epsilon^2)$ terms on the way,  and obtain the bicoherent-state path integral representation of (\ref{35}) as 
\be\label{pathintegral}
D^{({\theta})}(z_f,z_i;t) =  \int_{z(0)=z_i}^{z(t)=z_f} {\cal D}^2[z(t)] \, \exp\left[-{1\over 2}\int\limits_{0}^{t}\,dt\, (z^*(t)\dot z(t) - \dot z^*(t)z(t)) -i\int\limits_0^{t}\,dt\, H(z^*(t), z(t))\right]
 \en
where 
\be\label{classicalhamiltonian}
H(z^*(t), z(t))  = \omega_\theta (z^*z+\frac{1}{2}) 
\en
is the classical symbol for the hamiltonian (\ref{Hpseudo}) (with $B_\theta$ replaced by $z^*$ and $A_\theta$ by $z$), and where the measure ${\cal D}^2[z(t)]$ is the limit of the corresponding discrete measure $\left[\prod_{k=1}^{N-1} \,{d^2 z_k\over\pi} \right]$ in (\ref{41}) as $N\rightarrow\infty$. Integration in (\ref{pathintegral}) is carried over all paths $z(t)$ in the complex plane subjected to the boundary conditions $z(0)=z_i$ and $z(t) = z_f$, and each path  is weighed by the phase factor $e^{iS[z(t)]}$, where the (real-valued) classical action $S$ is 
\be\label{classicalaction}
S[z(t)] = {i\over 2}\int\limits_{0}^{t}\,dt\, (z^*(t)\dot z(t) - \dot z^*(t)z(t)) -\int\limits_0^{t}\,dt\, H(z^*(t), z(t))\,.
\en
As was commented upon already at the beginning of this section, the resulting bicoherent-state path integral (\ref{pathintegral}) is identical with the conventional coherent-state path integral of the hermitian oscillator \cite{schulman}.  The difference between the hermitian and nonhermitian systems will appear, of course, in position space matrix elements of propagators and in correlation functions.  

The semiclassical limit of (\ref{pathintegral}) is governed by the classical equations of motion resulting from (\ref{classicalaction}), namely, 
\be\label{classicaleom}
i\dot z = {\partial H\over\partial z^*} = \omega_\theta z\,,\quad i\dot z^* = -{\partial H\over\partial z} = -\omega_\theta z^*
\en
with the solution 
\be\label{solution}
z(t) = z_ie^{-i\omega_\theta t}\,,
\en
which we recognize as the argument of $\varphi^{(\theta)}$ in (\ref{34}).  This should come at no surprise, since coherent states in general are minimal uncertainly states.

 \newpage
 \section{The Feynman Path Integral}\label{Feynman}
We have derived in Section \ref{position-amplitude} the position matrix element  (\ref{meposition2})  of the propagator by transforming its bicoherent-state matrix element (\ref{35}) to the position basis. In this section we shall derive (\ref{meposition2}) directly from the Feynman path integral representation of the propagator. We shall do so by applying the Gelfand-Yaglom-Montroll method \cite{Gelfand-Yaglom, Montroll}, following Chapter 6 of \cite{schulman}. 

Before delving into Feynman path integration, let us present a quick way to compute this matrix element directly from that of the hermitian theory \eqref{meposition-hermitian}. To this end we invoke \eqref{similarity} and write 
\be\label{meposition3}
\langle x_f|e^{-iH_\theta t}|x_i\rangle  = \langle x_f|T_\theta e^{-ih_\theta t} T_\theta^{-1}|x_i\rangle  = \int\limits_{-\infty}^\infty dy_1 dy_2 \langle x_f|T_\theta |y_1\rangle \langle y_1|e^{-ih_\theta t} |y_2\rangle\langle y_2|T_\theta^{-1}|x_i\rangle 
\en
The matrix elements of $T_\theta$ and its inverse can be deduced from \eqref{squeeze-x} and the discussion following it. One finds
\begin{eqnarray}\label{squeeze-me}
 \langle x|T_\theta |y\rangle &=& e^{i\frac{\theta}{2}}\delta\left(e^{i\theta} x-y\right)\nonumber\\
 \langle y|T_\theta^{-1} |x\rangle &=&  \langle y|T_{-\theta} |x\rangle =e^{-i\frac{\theta}{2}}\delta\left(e^{-i\theta} y-x\right) = e^{i\frac{\theta}{2}}\delta\left(y-e^{i\theta}x\right) .
 \end{eqnarray}
For these expressions to make any sense, we have to think of $e^{\pm i\theta}$ as being real variables, because of the one-dimensional real Dirac delta functions\footnote{We stress once more: These genuinely real one-dimensional Dirac delta functions should not be confused with any complexified generalizations of delta functions which exist in the literature. See \cite{complexdelta1}-\cite{complexdelta3} and references therein.}. (In particular, we have used this assumption to arrive at the last equality in \eqref{squeeze-me}.) Effectively, we are taking the corresponding matrix element $ \langle x|T_{-iu} |y\rangle  =  \langle x| e^{i\frac{u}{2}(xp+px)} |y\rangle  = e^{\frac{u}{2}}\delta\left(e^{u} x-y\right)$ of the ordinary {\em unitary} squeezing operator (for real values of $u$) and its inverse, and then analytically continue to $u=i\theta$. Another way to interpret these expressions is to recall from Section \ref{Classical limit} that the classical variable $e^{i\theta}x$ corresponding to the observable $X_\theta$ is real.\\

Thus, by substituting \eqref{meposition-hermitian} and \eqref{squeeze-me} in \eqref{meposition3}, we obtain \eqref{meposition2} yet again. \\

Of course, we can express the propagator $ \langle y_1|e^{-ih_\theta t} |y_2\rangle$ of the hermitian oscillator in \eqref{meposition3}  as an ordinary Feynman path integral\,\cite{schulman}, which is the quickest way to write the Feynman path integral for  (\ref{meposition2}): These two path integrals are essentially the same, up to the action of the operators $T_{\pm\theta}$ at the end points. The paths to be summed over are just those of the ordinary hermitian oscillator, and they correspond to its position operator $x$, which is an observable in $\Hil$. It is of course similar to $X_\theta$, which is an observable in $\Hil_\theta$. Thus, there should be an equivalent derivation of this path integral by summing over paths in the real configuration space associated with the quasi-hermitian operator $X_\theta(t)$ discussed in Section \ref{nonhermitianxp}. The latter configuration space is identical to that of the hermitian problem, leading to the same path integral. 

To see this, we start by slicing the time segment as in Section \ref{slicingtime}, and write the propagator matrix element (\ref{meposition}) as 
$$\langle x_N| e^{-iH_\theta t}|x_0\rangle = \langle x_N\big| \left(e^{-iH_\theta \epsilon}\right)^N|x_0\rangle\,,$$ 
with endpoints $x_i=x_0$ at $t_0=0$ and $x_f = x_N$ at $t_N=t$. At each intermediate time $t_k$  we insert a resolution of unity 
\be\label{xk}
\int\limits_{-\infty}^\infty dx_k |x_k\rangle_{\!R}\,  {}_{L}\langle x_k| = \1
\en 
in terms of eigenstates of the operator $X_\theta(0)$ according to \eqref{tx} and \eqref{tx-completenessorthogonal} (evaluated at $t=0$), and write the matrix element as 
\begin{eqnarray}\label{position-slicing}
&& \langle x_N| e^{-iH_\theta t}|x_0\rangle = \int\limits_{-\infty}^\infty dx_{N-1}\cdots\int\limits_{-\infty}^\infty dx_1 \nonumber\\{}\nonumber\\
&&\langle x_N| e^{-iH_\theta (t_N-t_{N-1})} |x_{N-1}\rangle_{\!R}\,  {}_{L}\langle x_{N-1}|e^{-iH_\theta (t_{N-1}-t_{N-2})} |x_{N-2}\rangle_{\!R}\,  {}_{L}\langle x_{N-2}|\cdots \nonumber\\{}\nonumber\\
&&e^{-iH_\theta (t_{k+1}-t_{k})} |x_{k}\rangle_{\!R}\,  {}_{L}\langle x_{k}|e^{-iH_\theta (t_{k}-t_{k-1})}\cdots  |x_{2}\rangle_{\!R}\,  {}_{L}\langle x_{2}|e^{-iH_\theta (t_{2}-t_{1})} |x_{1}\rangle_{\!R}\,  {}_{L}\langle x_{1}|e^{-iH_\theta t_{1}} |x_{0}\rangle\,.\nonumber\\{}
\end{eqnarray}
Concentrate now on $x_k$ in (\ref{position-slicing}). Evidently, from (\ref{tx}) and (\ref{tx-completenessorthogonal}),  
\be\label{xk1}
\int\limits_{-\infty}^\infty dx_k\, e^{iH_\theta t_k}|x_k\rangle_{\!R}\,  {}_{L}\langle x_k|e^{-iH_\theta t_k} = \int\limits_{-\infty}^{\infty}dx_k\, |x_k;t\rangle_{\!R}\,  {}_{L}\!\langle x_;t|  = \1
\en
is just the resolution of unity in terms of complete biorthogonal set of eigenvectors of the quasi-hermitian operator $X_\theta(t_k)$. Therefore, 
\begin{eqnarray}\label{position-slicing1}
&& \langle x_N| e^{-iH_\theta t}|x_0\rangle = \int\limits_{-\infty}^\infty dx_{N-1}\cdots\int\limits_{-\infty}^\infty dx_1 \nonumber\\{}\nonumber\\
&&\langle x_N|e^{-iH_\theta t_N}|x_{N-1}; t_{N-1}\rangle_{\!R}\,  {}_{L}\langle x_{N-1}; t_{N-1} |x_{N-2}; t_{N-2}\rangle_{\!R}\,  {}_{L}\langle x_{N-2}; t_{N-2}|\cdots \nonumber\\{}\nonumber\\
&&|x_{k}; t_k\rangle_{\!R}\,  {}_{L}\langle x_k; t_k| \cdots  |x_{2}; t_2\rangle_{\!R}\,  {}_{L}\langle x_{2}; t_2 |x_{1}; t_1\rangle_{\!R}\,  {}_{L}\langle x_{1}; t_1 |x_{0}\rangle\,.
\end{eqnarray}
Thus, (\ref{position-slicing1}) expresses the propagation amplitude as a sum over all curves in the real configuration space associated with $X_\theta(t)$, which start at $x_0$ at $t_0=0$, terminate at $x_N$ at $t_N=t$, and pass through $x_1, x_2, \ldots x_{N-1}$ at the intermediate slicing points. The continuum limit $N\rightarrow\infty, N\epsilon=t$ of this sum, namely, Feynman's path integral for the propagator of $H_\theta$, is therefore a sum over all paths in the real $X_\theta(t)$-configuration space with the prescribed boundary conditions. 

Consider now the infinitesimal propagator between two consecutive intermediate points
\be\label{infi}
{}_{L}\langle x_{k+1}; t_{k+1} |x_{k}; t_k\rangle_{\!R} = \langle x_{k+1}; t_{k+1}|T_\theta^{-1} T_\theta |x_{k}; t_k\rangle =  \langle x_{k+1}; t_{k+1}|x_{k}; t_k\rangle\,,
\en
where we used \eqref{tx}. Not surprisingly, it reduces to the corresponding matrix element of the hermitian problem. Similarly, the matrix elements at the two endpoints are 
\begin{eqnarray}\label{endpoints}
\langle x_N|e^{-iH_\theta t_N}|x_{N-1}; t_{N-1}\rangle_{\!R} &=& \langle x_N | T_\theta e^{-ih_\theta \epsilon}|x_{N-1}\rangle\nonumber\\
{}_{L}\langle x_{1}; t_1 |x_{0}\rangle &=& \langle x_1; t_1|T_\theta^{-1}|x_0\rangle\,.
\end{eqnarray}
We end up simply with 
\begin{eqnarray}\label{position-slicing2}
&& \langle x_N| e^{-iH_\theta t}|x_0\rangle = \int\limits_{-\infty}^\infty dx_{N-1}\cdots\int\limits_{-\infty}^\infty dx_1 \nonumber\\{}\nonumber\\
&&\langle x_N | T_\theta e^{-ih_\theta \epsilon}|x_{N-1}\rangle\, \langle x_{N-1}; t_{N-1} |x_{N-2}; t_{N-2}\rangle\,  \langle x_{N-2}; t_{N-2}|\cdots \nonumber\\{}\nonumber\\
&&|x_{k}; t_k\rangle\,  \langle x_k; t_k| \cdots  |x_{2}; t_2\rangle\,  \langle x_{2}; t_2 |x_{1}; t_1\rangle\,  \langle x_1|e^{-ih_\theta\epsilon} T_\theta^{-1}|x_0\rangle\,,
\end{eqnarray}
which is  just a very cumbersome way of writing $ \langle x_N| (e^{-iH_\theta \epsilon})^N|x_0\rangle = \langle x_N| (T_\theta e^{-ih_\theta \epsilon} T_\theta^{-1})^N|x_0\rangle   = \langle x_N| T_\theta (e^{-ih_\theta \epsilon})^N T_\theta^{-1}  |x_0\rangle $, which leads us back to \eqref{meposition3}, as promised.

One can readily extend this discussion to computing matrix elements of time-ordered products $T (X_\theta(t_1)X_\theta(t_2)\cdots X_\theta(t_K))$ of the quasi-hermitian operator $X_\theta$ by means of Feynman path integration. We shall not pursue this issue any further here, as our main interest is consistency of our bicoherent and Feynman path ntegral results. We refer the interested reader to \cite{JR1,JR2} for further details on Feynman path integrals for generic quasi-hermitian systems, and the role of the metric $g$ therein. 

\subsection{Computing the path integral}
The limiting process discussed following (\ref{position-slicing1}) leads, in what by now is a well known procedure \cite{schulman}, to the path integral representation 
\be\label{Feynmanpath}
G(x_f,x_i;t)  =   \langle x_f| e^{-iH_\theta t}|x_i\rangle = \int\limits_{x(0)=x_i}^{x(t)=x_f}{\cal D}[x(\tau)]\, \exp\left(i\int\limits_0^t Ld\tau\right)\,,
\en
for the matrix element (\ref{meposition2}).  Here
\be\label{lagrangian}
L = \frac{1}{2\omega_\theta}\left(e^{i\theta}\dot x\right)^2  - \frac{\omega_\theta}{2}\left(e^{i\theta}x\right)^2 
\en
is the lagrangian corresponding to $H_\theta$ in (\ref{Htheta}), and summation is carried over all paths $x(t)$  subjected to the boundary conditions 
\be\label{bc}
x(0)=x_i\,,\quad x(t) = x_f\,.
\en 
As we shall see, this prescription takes correctly into account the effect of the operators $T_\theta$ and $T_\theta^{-1}$ at the endpoints in \eqref{meposition3}.
 
In order to proceed, we follow Chapter 6 of \cite{schulman}, and shift the integration variable in function space
\be\label{shift}
x(\tau) = x_{cl}(\tau) + \eta(\tau)\,,
\en
where $x_{cl}(\tau) $ is the solution of the classical equations of motion associated with the lagrangian (\ref{lagrangian}), subjected to (\ref{bc}). Thus, the functions $\eta(\tau)$ we integrate over are subjected to Dirichlet boundary conditions
\be\label{Dirichlet}
\eta(0)=\eta(t)=0\,.
\en
The jacobian for this shift of $x(t)$ by a fixed known function $x_{cl}(t)$ is of course just unity, that is, ${\cal D}[x(\tau)] = {\cal D}[\eta(\tau)]$. 

An overall multiplicative factor in the lagrangian, such as $e^{2i\theta}$, does not affect the Euler-Lagrange equations of motion, which therefore coincide with those for the lagrangian $L = \frac{1}{2\omega_\theta}\dot x^2  - \frac{\omega_\theta}{2}x^2 $ of the standard real harmonic oscillator with frequency $\omega_\theta$, namely,
\be\label{EL}
\ddot x + \omega_\theta^2 x = 0\,.
\en
The desired classical solution, subjected to (\ref{bc}), is readily found to be 
\be\label{xcl}
x_{cl}(\tau) = x_i {\sin(\omega_\theta(t-\tau))\over\sin(\omega_\theta t)}  + x_f {\sin(\omega_\theta\tau)\over\sin(\omega_\theta t)} \,.
\en 
We now substitute (\ref{shift}) and (\ref{xcl}) in (\ref{lagrangian}) and compute the corresponding action functional
\begin{eqnarray}\label{action}
S[x(\tau)] &=& S[x_{cl}(\tau)] + \frac{e^{2i\theta}}{2}\int\limits_0^t \left(\frac{1}{\omega_\theta}\dot\eta^2 - \omega_\theta\eta^2\right)\,d\tau\nonumber\\{}\nonumber\\
&=& S[x_{cl}(\tau)] - \frac{e^{2i\theta}}{2\omega_\theta}\int\limits_0^t \eta\left(\frac{d^2}{d\tau^2} + \omega^2_\theta\right)\eta\,d\tau\,,
\end{eqnarray}
where in the last integral we integrated by parts and used the boundary conditions (\ref{Dirichlet}). Note that in (\ref{action}) there is no term linear in $\eta$, because $x_{cl}(t)$ is an extremal configuration of the action. After some additional work we obtain the classical action as 
\be\label{Sclass}
S[x_{cl}(\tau)] = \int\limits_0^t \, L(x_{cl},\dot x_{cl})\,d\tau = \frac{e^{2i\theta}}{2\sin\omega_\theta t}\left[(x_i^2+x_f^2)\cos\omega_\theta t - 2x_ix_f\right]\,.
\en
As a consistency check of the latter expression, note that for very small $t$ it tends to $$\frac{e^{2i\theta}}{2\omega_\theta}\frac{(x_f-x_i)^2}{t}\,,$$
namely, the action of a free particle of mass $\frac{e^{2i\theta}}{\omega_\theta}$ moving from $x_i$ to $x_f$ in time $t$, consistent with the kinetic term of (\ref{lagrangian}). 
Next, substitute (\ref{action}) and (\ref{Sclass}) in (\ref{Feynmanpath}) and write
\be\label{Feynmanpath1}
G(x_f,x_i;t)  = e^{iS[x_{cl}]}\int\limits_{\eta(0)=0}^{\eta(t)=0}{\cal D}[\eta(\tau)]\, \exp\left[-i\frac{e^{2i\theta}}{2\omega_\theta}\int\limits_0^t \eta\left(\frac{d^2}{d\tau^2} + \omega^2_\theta\right)\eta\,d\tau\right]\,.
\en
The last Gaussian functional integration can be carried out explicitly by employing the Gelfand-Yaglom-Montroll method \cite{Gelfand-Yaglom, Montroll}, as explained in \cite{schulman}. To this end we first solve the {\em initial value} problem 
\begin{eqnarray}\label{initialvalue}
&&\frac{d^2f(\tau)}{d\tau^2} +\omega^2_\theta f(\tau) = 0\nonumber\\
&&f(0)=0\,,\quad \dot f(0) = 1
\end{eqnarray}
associated with the Dirichlet Sturm-Liouville operator $\frac{d^2}{d\tau^2} + \omega^2_\theta$ in (\ref{Feynmanpath1}). The solution is 
\be\label{ftau}
f(\tau;\omega_\theta) = \frac{\sin(\omega_\theta\tau)}{\omega_\theta}\,.
\en
Then, one can show that 
\be\label{Gaussian}
\int\limits_{\eta(0)=0}^{\eta(t)=0}{\cal D}[\eta(\tau)]\, \exp\left[-i\frac{e^{2i\theta}}{2\omega_\theta}\int\limits_0^t \eta\left(\frac{d^2}{d\tau^2} + \omega^2_\theta\right)\eta\,d\tau\right] = \left({e^{2i\theta}/\omega_\theta\over 2\pi i f(t)}\right)^{\frac{1}{2}} = {e^{i\theta}\over\sqrt{2\pi i\sin\omega_\theta t}}\,,
\en
because $f(t;\omega_\theta)$ is essentially the finite regularized form of the functional determinant of the operator $\frac{d^2}{d\tau^2} + \omega^2_\theta$ appearing in the Gaussian integral. 

Finally, we substitute (\ref{Sclass}) and (\ref{Gaussian}) in (\ref{Feynmanpath1}), and obtain our desired result
\be\label{FeynmanFinal}
\langle x_f |e^{-iH_\theta t}|x_i\rangle = {e^{i\theta}\over\sqrt{2\pi i\sin\omega_\theta t}}\,\exp\left\{ \frac{ie^{2i\theta}}{2\sin\omega_\theta t}\left[(x_i^2+x_f^2)\cos\omega_\theta t - 2x_ix_f\right] \right\}\,,
\en 
which coincides with (\ref{meposition2}). Consistency of (\ref{meposition2}), which was derived from the bicoherent-state path integral, and (\ref{FeynmanFinal}), which was derived directly from Feynman's path integral, vindicates our novel bicoherent-state path integral derivation. 

\section{Conclusions and outlook}\label{conclusions}

In this paper we have introduced, for the first time, bicoherent-state path integration as a method for quantizing non-hermitian systems. We have applied it to the concrete and very simple system given by (\ref{200}), and used it to obtain its propagator. We have verified the consistency of our bicoherent-state results with direct Feynman path integration. On the way, we have elucidated the type of paths summed over in the Feynman path integral.

The present work opens the way to applying bicoherent-state path integration to more interesting quasi-hermitian systems which consist of many interacting degrees of freedom, such as quasi-hermitian quantum field theories or statistical mechanical systems. As one final comment, we just mention that our results for the {\em Feynman} path integral can be extended also to non-hermitian systems which are not quasi-hermitian, as long as the non-hermitian position or field operators are diagonalizable in terms of biorthogonal bases. 


\newpage
\setcounter{equation}{0}
\setcounter{section}{0}
\renewcommand{\theequation}{A.\arabic{equation}}
\renewcommand{\thesection}{Appendix:}
\section{The Gaussian integral (\ref{Iint})}\label{app}
\vskip 5mm
\setcounter{section}{0}
\renewcommand{\thesection}{A}

The bilinear form 
\be\label{bilibear}
\xi = |z_f|^2+|z_i|^2+\frac{1}{2}(z_f^2 + z_i^{*2}) - \sqrt{2}e^{i\theta}(z_f x_f +z^*_i x_i)  -z_f^*z_ie^{-i\omega_\theta t}
\en
in the exponential of the Gaussian integral (\ref{Iint}) may be written in matrix form as 
\be\label{bilinear2}
\xi = \frac{1}{2} {\bf u}^T {\bf M} {\bf u} - \frac{1}{\sqrt{2}}\left({\bf u}^T{\bf v} + {\bf v}^T{\bf u} \right)\,,
\en
with 
\be\label{uv}
{\bf u} = \left(\begin{array}{c} {\rm Re}z_f\\ {\rm Im}z_f\\ {\rm Re}z_i\\ {\rm Im}z_i\end{array}\right)\,,\quad\quad {\bf v} = \left(\begin{array}{c} e^{i\theta}x_f\\ ie^{i\theta}x_f\\ e^{i\theta}x_i\\ -ie^{i\theta}x_i\end{array}\right)
\en
and with the symmetric matrix
\be\label{Mmatrix}
{\bf M} = \left(\begin{array}{cccc} 3  &  i  & -e^{-i\omega_\theta t}  & -ie^{-i\omega_\theta t}\\  i & 1 & ie^{-i\omega_\theta t}  & -e^{-i\omega_\theta t}\\ -e^{-i\omega_\theta t} &  ie^{-i\omega_\theta t}  & 3 & -i\\  -ie^{-i\omega_\theta t}  & -e^{-i\omega_\theta t} & -i & 1\end{array}\right)
\en
In these notations, the integral (\ref{Iint}) may be written as
\be\label{Iint1}
I(x_f,x_i;t) = \int_{\mathbb{R}^4} {d^4{\bf u}\over\pi^2} e^{-\xi}\,.
\en
The eigenvalues of ${\bf M}$ are $2(1\pm e^{-i\omega_\theta t/2})$ and $2(1\pm ie^{-i\omega_\theta t/2})$, all with {\em non-negative} real parts. Thus, (\ref{Iint}) converges (for most values of $t$), and it can be computed in the standard way, by completing the squares in $\xi$. The result is 
\be\label{Iint2}
I(x_f,x_i;t) = 4\, {e^{{\bf v}^T {\bf M}^{-1} \,{\bf v}}\over \sqrt{\det{\bf M}}}\,.
\en
One can compute 
\be\label{det}
\det{\bf M} = 32\, i e^{-i\omega_\theta t}\sin\omega_\theta t
\en
and 
\be\label{invM}
{\bf M}^{-1} = \frac{1}{4(1-e^{-2i\omega_\theta t})}\left(\begin{array}{cccc} 1-e^{-2i\omega_\theta t}  &  -i(1-e^{-2i\omega_\theta t})  & 0  & 0\\  -i(1-e^{-2i\omega_\theta t}) & 3+e^{-2i\omega_\theta t}  & 0  & 4\,e^{-i\omega_\theta t}\\ 0 & 0  & 1-e^{-2i\omega_\theta t}  & i(1-e^{-2i\omega_\theta t}) \\  0  & 4\,e^{-i\omega_\theta t} & i(1-e^{-2i\omega_\theta t})  & 3+e^{-2i\omega_\theta t}\end{array}\right)\,.
\en
Putting everything together we obtain 
\be\label{Iintfinal}
I(x_f,x_i;t) = {e^{\frac{i\omega_\theta t}{2}}\over\sqrt{2i \sin\omega_\theta t}}  \exp\left\{\frac{e^{2i\theta}}{1 - e^{2i\omega_\theta t}}\left(x_f^2 + x_i^2 - 2e^{i\omega_\theta t}x_fx_i\right)\right\}\,.
\en
Finally, substitution of (\ref{Iintfinal}) in (\ref{meposition1}) leads to (\ref{meposition2}).

\section*{Acknowledgements}

This collaboration was initiated when the authors met at the International Centre for Theoretical Sciences (ICTS) during a visit for participating in the program `Non-Hermitian Physics - PHHQP XVIII' (Code: ICTS/nhp2018/06).
This work was partially supported by the University of Palermo and by the Gruppo Nazionale di Fisica Matematica of Indam, and by the Israel Science Foundation (grant No.\,2040/17). J.F. also thanks the University of Palermo for financial support via CORI.

\end{document}